\documentclass{article}
\usepackage{spconf,amsmath,graphicx}
\usepackage{bbm}
\usepackage{amsfonts,amsthm} 
\usepackage{bm}
\usepackage{cite}
\usepackage{xcolor}

\newcommand{\brho}{\boldsymbol{\rho}}
\newcommand{\bgamma}{\boldsymbol{\gamma}}
\newcommand{\bix}{\bm{x}}
\newcommand{\bx}{\mathbf{x}}
\newcommand{\bd}{\mathbf{d}}
\newcommand{\bw}{\mathbf{w}}
\newcommand{\bu}{\mathbf{u}}
\newcommand{\bz}{\mathbf{z}}
\newcommand{\ba}{\mathbf{a}}

\newcommand{\bP}{\mathbf{P}}
\newcommand{\bX}{\mathbf{X}}
\newcommand{\calF}{\mathcal{F}}
\newcommand{\calD}{\mathcal{D}}
\newcommand{\calI}{\mathcal{I}}
\newcommand{\calJ}{\mathcal{J}}
\newcommand{\calR}{\mathcal{R}}
\newcommand{\bbR}{\mathbb{R}}
\newcommand{\bbC}{\mathbb{C}}

\newcommand{\bbmR}{\mathbbm{R}}

\DeclareMathOperator*{\argmin}{argmin}

\title{DEEP DENOISING PRIOR-BASED SPECTRAL ESTIMATION FOR PHASELESS SYNTHETIC APERTURE RADAR}
\name{S. Kazemi, B. Yonel, B. Yazici\thanks{This work was supported in part by the Air Force Office of Scientific Research (AFOSR) under agreement FA9550-19-1-0284, in part by Office of Naval Research (ONR) under agreement N00014-18-1-2068, in part by the National Science Foundation (NSF) under Grant No ECCS-1809234 and in part by the United States Naval Research Laboratory (NRL) under agreement N00173-21-1-G007.}}
\address{Rensselaer Polytechnic Institute \\
	Department of Electrical, Computer, and Systems Engineering \\
	110 8th St, Troy, NY 12180}

\begin{document}
%
\maketitle
\begin{abstract}
Incoherent processing for synthetic aperture radar (SAR) is a promising approach that enables low implementation costs, simplified hardware designs and operations in high frequency spectrum compared to the conventional imaging methods using coherent processing.
Existing non-convex phaseless imaging algorithms offer recovery guarantees over limited range of forward models.
In recent years, several deep learning (DL) based techniques have been introduced with the goal of extending applicability of phaseless imaging techniques to wave-based imaging modalities by addressing fundamental challenges, such as, lack of redundancy, non-uniqueness issues encountered commonly with inverse scattering models.
In this paper, we introduce a DL-based phaseless SAR imaging approach that is designed under the premise that the spectral estimation technique, widely used for initializing non-convex phase retrieval algorithms, has significance far beyond generating good initial points.
We extend the iterative power method for spectral estimation by using deep denoisers at each iteration, and subsequently design a deep imaging network within the plug-and-play framework.
Finally, we verify the feasibility of our approach using synthetic SAR measurements.
\end{abstract}
\begin{keywords}
Synthetic aperture radar, phase retrieval, phaseless imaging, deep learning
\end{keywords}
\section{Introduction}
\label{sec:intro}
Imaging an unknown quantity of interest from the magnitude or intensity values of its linear mapping has gained significant interest in applications to wave-based imaging using electromagnetic and acoustic waves
~\cite{moscoso2019synthetic, chen2017phaseless}.
However, the deployment of phaseless synthetic aperture radar (SAR) has been quite limited in practice.
For SAR imaging at high frequencies,
it is challenging to accurately measure the phase information
~\cite{novikov2015illumination}.
Phase errors are commonplace
due to the presence of phase modulator noise, and temperature variations~\cite{zhang2009inverse}.
Moreover, while imaging using wide apertures, maintaining coherence across phase measurements becomes demanding, requiring complex and sophisticated receiver designs.
Additionally, random fluctuation in the transmission medium properties and uncompensated platform trajectory deviations also lead to phase errors by giving rise to inaccurate model assumptions.
Phaseless SAR, on the other hand,
only measures the magnitude or intensity values of back-scattered waves, hence enables simple hardware designs and provides robustness to phase errors that can occur during data acquisition.

\vspace{-1pt}
Despite potential benefits, phaseless SAR imaging encounters a number of key challenges in designing algorithms with recovery guarantees.
The quadratic nature of the problem implies that the true unknown is only recoverable up to certain trivial ambiguities~\cite{candes2015phase_IEEE, sun2018geometric, yonel2020deterministic}.
Moreover, for exact recovery, the loss of phase information needs to be compensated either during the acquisition phase via hardware modifications~\cite{novikov2015illumination} or during imaging by advanced algorithms~\cite{soltanolkotabi2019structured, yuan2019_sparseWF, Xue2022_nonconvex_denoising}.
Acquisition level compensations lead to using high number of measurements, and designing optimal illumination strategies~\cite{vaswani2020nonconvex, soltanolkotabi2019structured, novikov2015illumination}.
However, collecting prohibitively large number of measurements
from spatially distributed transceivers as well as manipulating sampling vectors via hardware modifications are not desirable in practice.
In this paper, we instead adopt an algorithmic approach by utilizing prior information about the target class of interest.
Exact recovery guarantees of non-convex phase retrieval algorithms~\cite{candes2015phase_IEEE, zhang2017RWF} rely on good initial estimates that reside within small and locally strictly convex neighborhoods of the unknown of interest.
This can be accomplished under certain conditions by spectral estimation methods~\cite{monardo2019sensitivity}, which use the leading eigenvector of back-projected intensity measurements, as well as its variants~\cite{zhang2017RWF, yuan2019_sparseWF}.
Sampling vectors for SAR typically do not possess orthogonality or desirable statistical properties, such that, the back-projected intensity measurements are sufficiently accurate with respect to the underlying Kronecker scene~\cite{mason2015passive}, especially at low number of measurements to number of unknown ratios.

In recent years, deep learning (DL) based phaseless imaging methods that incorporate prior information with first order iterative algorithms~\cite{hand2018phase, kazemi2022unrolled} and with the power method for spectral estimation~\cite{liu2022generative} have been introduced.
A challenge for phaseless SAR is the non-uniqueness issue in the far field due to the translation invariance of the associated forward maps~\cite{zhang2017recovering_farfield}.
These DL based methods, which implement generative, decoding or denoising type priors~\cite{hand2018phase, kazemi2022unrolled}, restrict the estimated images to specific manifolds learned from training data.
These priors are particularly suitable for phaseless SAR as the learned image manifolds can be optimized to eliminate undesirable translation replicas.
In~\cite{liu2022generative}, prior information is imposed while learning the leading eigenvector of the back-projected intensity measurements by projecting onto the range of a deep variational auto-encoder.
However, theoretical analysis relies on i.i.d. Gaussian distributed sampling vectors that are unsuitable for SAR.
In~\cite{kazemi2022interferometric}, we introduced a deep denoising prior-based spectral estimation approach for SAR interferometric imaging problem using the plug-and-play (PnP)~\cite{kamilov2017plug_nonlinear} and regularization by denoising (RED)~\cite{romano2017little} frameworks.
We designed end-to-end imaging networks via unrolling~\cite{naimipour2020_upr}, and hence enabled a supervised denoiser training scheme that uses both the ground truth sample images and the corresponding cross-correlated measurements to attain model aware denoising operators.

In this paper, we present a PnP and spectral estimation based imaging method for phaseless SAR by extending our earlier approach in~\cite{kazemi2022interferometric}.
Our unrolled imaging network and the subsequent supervised denoiser training scheme is motivated by analyzing the role that an optimal regularization term associated with the denoiser should play towards attaining improved reconstructions compared to the spectral estimation method.
We verified the feasibility of our proposed approach by using simulated SAR phaseless measurements and compared performance with the Wirtinger flow (WF) algorithm for phaseless imaging~\cite{candes2015phase_IEEE, yonel2020deterministic}.

\section{Problem Statement}
Let $\bx = (x_1, x_2, x_3)\in\bbR^3$ and $\bix = (x_1, x_2)\in\bbR^2$ be the location variables in $3$-D and $2$-D Cartesian coordinate systems, respectively.
For ground topography $\phi:\bbR^2\mapsto\bbmR$, location $\bx$ on the ground surface is represented as $(\bix, \phi(\bix))$.
We denote the temporal frequencies by $\omega$ and the slow-time points by $s$.
The transmitting and receiving sensor locations at the various slow-time points are $\bgamma_T(s)\in\bbR^3$ and $\bgamma_R(s)\in\bbR^3$, respectively.
We assume a homogeneous background medium with electromagnetic wave speed $c_0$.
Let $\rho:\bbR^2\mapsto\bbmR$ denote the unknown reflectivity function for the points on the ground topography .
We further assume that the target dimensions, scene reflectivities, and the transmission bandwidths are such that, the Born approximation related model simplification is valid.
Then, under the start-stop signal model, SAR received signal $f(\omega, s)$ relates to $\rho$ as follows:
\begin{align}
\label{eq:sar_data_with_phase} f(\omega, s) & = \int\exp\left(-\frac{\text{i}}{c_0}\omega\phi(s, \bix)\right)A(\omega, s, \bix)\rho(\bix)d\bix.
\end{align}
Phase term $\phi(s, \bix)$ refers to the total distance $|\bgamma_T(s) - \bix| + |\bix - \gamma_R(s)|$ travelled by the reflected signal at slow-time $s$.
Amplitude $A(\omega, s, \bix)$ relates to the fast-time frequencies, beam-patterns of the transmitting and receiving antennas, and geometric spreading factor~\cite{mason2015passive}, and it is assumed to be constant-valued here for simplification.

For phaseless SAR imaging, only the intensity value of $f(\omega, s)$ is relevant which we denote by $d(\omega, s)$.
From~\eqref{eq:sar_data_with_phase},
\begin{align}
\label{eq:sar_data_intensity} d(\omega, s) & = \int\text{e}^{-\frac{\text{i}}{c_0}\omega(\phi(s, \bix) - \phi(s, \bix'))}\rho(\bix)\overline{\rho(\bix')}d\bix d\bix'.
\end{align}
Phase term $\phi(s, \bix) - \phi(s, \bix')$ 
can be simplified by using the small scene and far field assumptions and the Taylor series expansion, which leads to
\begin{align}
\label{eq:sar_data_intensity_ss_ff} d(\omega, s) & = \int\text{e}^{-\frac{\text{i}\omega}{c_0}(\hat{\bgamma}_T(s) + \hat{\bgamma}_R(s)).(\bix' - \bix)}\rho(\bix)\overline{\rho(\bix')}d\bix d\bix'.
\end{align}
$\hat{(.)}$ notation over a vector is used to represent the unit vector along its direction.

In this paper, we propose an algorithm for recovering the unknown ground reflectivities at a finite number of locations.
To this end, we discretize the scene into $N$ locations at $\{\bx_n\}_{n = 1}^N$, and define a corresponding unknown or ground truth image vector $\brho^*\in\bbC^N$ whose $n^{th}$ component is calculated as $\rho(\bx_n)$.
We also assume that $M$ total measurements are collected by sampling over the transmission bandwidth using $K$ equi-distant points, $\{\omega_k\}_{k = 1}^K$, at each of the $S$ slow time samples, i.e., $M = SK$.
Let $\bP^*$ be the Kronecker scene for image vector $\brho^*$, i.e., $\bP^* = \brho^*{\brho^*}^H$, and $\ba_m\in\bbC^N$ be the $m^{th}$ sampling vector whose $n^{th}$ element $\ba_m(n)$ is calculated as $\text{e}^{\frac{-\text{i}\omega}{c_0}(\hat{\bgamma}_T(s) + \hat{\bgamma}_R(s)).\bix_n}$.
We represent the $m^{th}$ intensity measurement by $d_m$, where $m = (s - 1)K + k$, $s\in[S]$ and $k\in[K]$.
Then, from~\eqref{eq:sar_data_intensity_ss_ff}, we observe that $d_m$ can be expressed as $\ba^H_m\bP^*\ba_m$.
Let $\bd\in\bbR^M_+$ be the intensity measurement vector composed of $\{d_m\}_{m = 1}^M$, and $\calF:\bbC^{N \times N}\mapsto\bbC^M$ be a known lifted forward map such that 
\begin{align}
\label{eq:data_model_discrete} \bd & = \calF(\bP^*).
\end{align}
Our objective is to estimate the unknown image $\brho^*$, up to a global phase factor, by using the known values of $\bd$ and $\calF$.

\section{Proposed Approach}
\subsection{Spectral Estimation}
Non-convex phaseless imaging methods employ spectral estimation or its variants for initialization, and estimate directly at the signal level, instead of calculating a lifted domain reconstruction from the linear data model in~\eqref{eq:data_model_discrete}.
These methods are commonly equipped with theoretical recovery guarantees for sampling vectors with ideal probability distributions~\cite{candes2015phase_IEEE, zhang2017RWF}, as well as for limited deterministic cases~\cite{yonel2020deterministic}.
Spectral estimation sets the initial image $\brho_0\in\bbC^N$ as $\sqrt{\lambda_0}\bu_1$, where $\lambda_0 = \frac{1}{\sqrt{2M}}\|\bd\|$~\cite{yonel2020deterministic},
and $\bu_1\in\bbC^N$ represents the leading eigenvector of the spectral matrix $\hat{\bX}\in\bbC^{N \times N}$, which is
calculated as
\begin{align}
\hat{\bX} & = \frac{1}{M}\calF^H\bd.
\end{align}
For phaseless SAR at low $M/N$ ratios, $\frac{1}{M}\calF^H\calF$ may deviate significantly from the ideal identity operator.
To this end, we observe from our approach in~\cite{kazemi2022interferometric} that by applying a set of optimally learned denoising operators, while estimating the leading eigenvector of the spectral matrix using the power method, faster convergence rate and improved reconstruction qualities are attained compared to using hand-crafted priors, as well as when spectral estimation is followed by gradient descent updates for further refinements.

\subsection{Deep Network for Phaseless SAR imaging}
\begin{figure*}[ht]
\centering
\includegraphics[width=11cm]{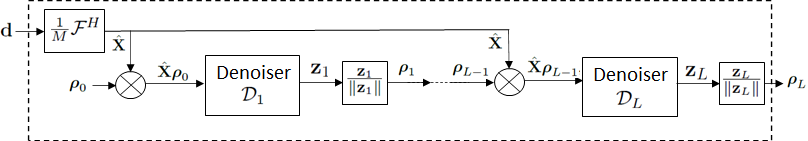}
\caption{Imaging network for phaseless SAR.}
\label{fig:network_model_phaseless_SAR}
\end{figure*}
In this paper, we extend our approach in~\cite{kazemi2022interferometric} to the phaseless SAR imaging problem.
Unlike the generative prior based techniques in~\cite{hand2018phase, kazemi2022unrolled}, our proposed DL-based method estimates the unknown image directly, instead of determining its encoded representation, by modifying the manifold over which the leading eigenvector of $\hat{\bX}$ is searched for by using learned prior information.
This is accomplished by applying a denoiser after each power method update to restrict the estimated image to the range of the denoise similar to the PnP framework,
and we model the denoiser by a deep network (DN) with trainable parameters.

Our proposed algorithm estimates the $\ell_2$ normalize ground truth image vector, which we denote by $\brho^*_n\in\bbC^N$, i.e., $\brho^*_n = \brho^*/\|\brho^*\|$.
We represent the estimated quantity by $\hat{\brho}_n\in\bbC^N$.
The algorithm starts from a fixed initial vector $\brho_0\in\bbC^N$, which is followed by iterative update steps.
At the $l^{th}$ iteration, the updated image is calculated as follows:
\begin{align}
\label{eq:power_method_update} \bw_l & = \hat{\bX}\brho_{l - 1}, \\
\label{eq:denoising} \bz_l & = \calD_l(\bw_l), \\
\label{eq:normalization} \brho_l & = \bz_l / \|\bz_l\|.
\end{align}
$\calD_l:\bbC^N\mapsto\bbC^N$ represents the denoising operator applied at the $l^{th}$ iteration.
When same denoising operator, $\calD:\bbC^N\mapsto\bbC^N$, is applied at each update, we can interpret it as the proximal operator associated with an implicit regularization term $\calR:\bbC^N\mapsto\bbC^N$, i.e.,
\begin{align}
\label{eq:calD_calR} \calD(\bw_l) = \argmin_{\bz\in\bbC^N} \|\bz - \bw_l\|^2 + \calR(\bz).
\end{align}
PnP framework directly applies $\calD$ without explicitly determining a corresponding $\calR$ term~\cite{kamilov2017plug_nonlinear}.
DL-based implementations model $\calD$ by a DN, whose optimal parameter values are learned using a set of noisy training images and the corresponding ground truth noise-free ones without considering the data model for the inversion task at hand.
When applied to ill-posed inversion tasks, such as, imaging for phaseless SAR under low sample complexities, such model independent denoiser designs may not be sufficient to generate good reconstructions.

To illuminate this point further in the context of the spectral estimation task for phaseless SAR, we next state the optimization problem that our algorithm in~\eqref{eq:power_method_update} to~\eqref{eq:normalization} addresses when $\calD_l$ is replaced by $\calD$: 
\begin{align}
\label{eq:optimization_prob} \hat{\brho}_n & = \argmin_{\brho\in\bbC^N} \calJ_S(\brho) + \calI_n(\brho) + \calR(\brho).
\end{align}
$\calI_n$ denotes an indicator whose value is large if the $\ell_2$ norm of its argument is not $1$.
Data fidelity term $\calJ_S(\brho)$ is defined as follows:
\begin{align}
\calJ_S(\brho) & = -\brho^H\hat{\bX}\brho + \|\brho\|^2.
\end{align}
By expanding $\calJ_S$,
it is straight-forward to verify that
\begin{align}
\label{eq:calJ_S_expansion} \calJ_S(\brho) = - |\brho^H\brho^*|^2 + \|\brho\|^2 - \brho^H\delta(\brho^*{\brho^*}^H)\brho,
\end{align}
where
\begin{align}
\delta = \frac{1}{M}\calF^H\calF - \calI.
\end{align}
This indicates that an ideal regularization term, imposed indirectly by the denoiser, should reduce the contribution of the undesirable third term in~\eqref{eq:calJ_S_expansion} for all possible $\brho^*$ from the image class of interest.
This term captures the perturbation present in the spectral matrix from the ideal one, $\bP^*$, due to the deviation of $\frac{1}{M}\calF^H\calF$ operator from identity.
For SAR phaseless imaging with limited number of slow-time points, this perturbation term can become significant, and this explains the resulting poor spectral estimation quality when no prior information is utilized.

The dependence of the ideal regularization term, i.e. $\brho^H\delta(\brho^*{\brho^*}^H)\brho$, on the unknown ground truth images makes it challenging to apply it directly by deriving an associated proximal operator.
Instead we implement it indirectly by using DL-based modelling of the denoisers at the various iterations of our algorithm.
Additionally, to facilitate model-aware denoiser training, we implement the unrolling technique~\cite{naimipour2020_upr}.
Unrolling refers to the mapping of $L$ iterations of an algorithm into $L$ stages of a DN.
The estimated image $\hat{\brho}_n$ is then set to $\brho_L$.
This results in an end-to-end imaging network with input $\bd$ and output $\brho_L$, and hence it can be trained in a supervised manner by using a set of ground truth training images $\{\brho^*_t\}_{t = 1}^T$ and the associated intensity measurement vector $\{\bd_t\}_{t = 1}^T$.
Our overall phaseless SAR imaging network is shown in Fig.~\ref{fig:network_model_phaseless_SAR}.

\section{Numerical Simulations}
%
\begin{figure}[htb]
  \centering
  \includegraphics[width=3.5cm]{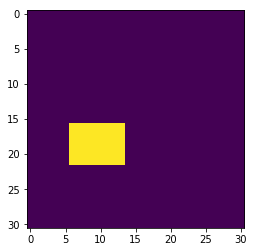}
  \caption{Ground truth image with a rectangular object.}
  \label{fig:ground_truth}
\end{figure}

\begin{figure}[htb]
\centering
%
\begin{minipage}[b]{0.45\linewidth}
  \centering
  \centerline{\includegraphics[width=3.5cm]{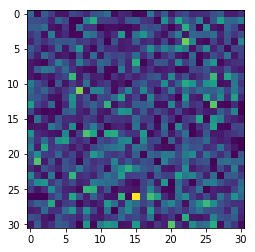}}
  \centerline{(b) WF (5dB)}
\end{minipage}
\begin{minipage}[b]{0.45\linewidth}
  \centering
  \centerline{\includegraphics[width=3.5cm]{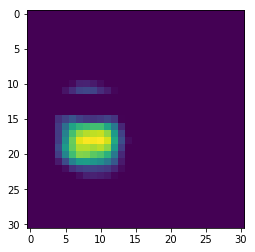}}
  \centerline{(c) Proposed approach (5dB)}
\end{minipage}
\\
\begin{minipage}[b]{0.45\linewidth}
  \centering
  \centerline{\includegraphics[width=3.5cm]{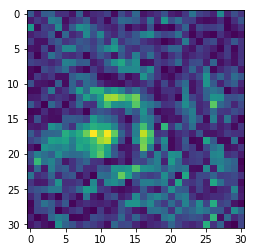}}
  \centerline{(b) WF (10dB)}
\end{minipage}
\begin{minipage}[b]{0.45\linewidth}
  \centering
  \centerline{\includegraphics[width=3.5cm]{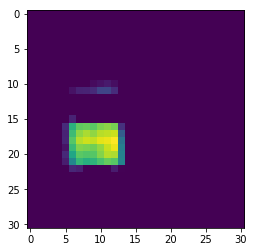}}
  \centerline{(c) Proposed approach (10dB)}
\end{minipage}
\caption{Reconstructed images using the WF algorithm with $150$ iterations and the proposed approach for intensity measurements with $5$ dB and $10$ dB additive noise.}
\label{fig:reconstructed_scenes}
\end{figure}
In this section, we verified the feasibility of our approach by using simulated SAR scenes and intensity measurements.
We considered a dataset of images, each with a single extended and randomly located rectangular target of random dimensions.
Our training set is composed of $5000$ samples, each having the ground truth image vector and the corresponding SAR intensity measurements available, while the test set contains $50$ samples, each having only the intensity measurements information.
For each sample, we assume that the area within the field of view of the SAR, that contains the rectangular target, has dimensions $62$ m $\times 62$ m.
The SAR is mono-static and operates in the spot-light mode with its moving platform having a circular trajectory of radius $10$ km, aperture length $\pi$ radian and an altitude of $7$ km.
The central frequency and the transmission bandwidth are $9.9$ GHz and $75$ MHz, respectively.
We reconstruct the scenes from each set of measurements as $31 \times 31$ pixels images.
The $M/N$ ratio used is $2$ with the number of slow time points, distributed at equi-distant locations over the aperture, being $62$ and the number of frequency samples at each slow time point being $31$.
We additionally considered that additive Gaussian distributed noise of $5$ dB and $10$ dB are present on the measured intensity values of the scattered signals.

We implemented the denoisers by using convolutional neural networks (CNN) with $16$ convolution layers and $\text{relu}(.)$ type nonlinear
activation functions.
We used $4$ iterations for constructing our unrolled imaging network from Fig.~\ref{fig:network_model_phaseless_SAR}.
Also, to limit the total number of trainable parameters, we gradually increased this value and used the same set of denoiser parameter values for two consecutive layers.
For the example ground truth image in Fig.~\ref{fig:ground_truth}, the reconstructed images are shown in Fig.~\ref{fig:reconstructed_scenes} using our proposed approach, along with the ones obtained using the WF algorithm~\cite{candes2015phase_IEEE, yonel2020deterministic} for measurements with two different noise levels.
For the given $M/N$ ratio, our approach out-performs the WF algorithm in terms of mean squared error (MSE) metric at both SNR levels.
Average computation times per test sample using our approach and the WF algorithm to attain these MSE levels are approximately $0.004$ seconds and $21.5$ seconds, respectively, indicating an improvement in computation time by our approach.

\section{Conclusion}
In this paper, we presented a DL-based SAR phaseless imaging method by modifying the spectral estimation approach using learned prior information.
Our approach is empirically observed to out-perform the iterative non-DL based WF algorithm using a simulated SAR dataset.
Theoretically studying sufficient conditions on the forward map and the denoiser architecture for guaranteeing exact recovery, as well as, its implementation using complex scenes and real SAR measurements are two important open problems that we leave for future work.


\newpage

\bibliographystyle{IEEEbib}
\bibliography{strings, refs}

\end{document}